# The Climates of Other Worlds: A Review of the Emerging Field of Exoplanet Climatology[*]

Aomawa L. Shields
Department of Physics and Astronomy, University of California, Irvine, 4129 Frederick Reines Hall, Irvine, CA 92697-4575, USA; shields@uci.edu

## Abstract

The discovery of planets orbiting stars other than the Sun has accelerated over the past decade, and this trend will continue as new space- and ground-based observatories employ next-generation instrumentation to search the skies for habitable worlds. However, many factors and processes can affect planetary habitability and must be understood to accurately determine a planet's habitability potential. While climate models have long been used to understand and predict climate and weather patterns on the Earth, a growing community of researchers has begun to apply these models to extrasolar planets. This work has provided a better understanding of how orbital, surface, and atmospheric properties affect planetary climate and habitability; how these climatic effects might change for different stellar and planetary environments; and how the habitability and observational signatures of newly discovered planets might be influenced by these climatic factors. This review summarizes the origins and evolution of the burgeoning field of exoplanet climatology, discusses recent work using a hierarchy of computer models to identify those planets most capable of supporting life, and offers a glimpse into future directions of this quickly evolving subfield of exoplanet science.

## 1. Introduction

Where once it was believed that the Earth was the only planet capable of hosting life, we now know of dozens of planets with the potential to sustain the requirements for life as we know it to thrive, among the thousands of extrasolar planets identified. This reality takes us far and away from that initial discovery in 1995 of the first planet orbiting another main-sequence star, 51 Pegasi b (Mayor & Queloz 1995; Marcy et al. 1997). 51 Peg b was found using the radial velocity technique, which measures changes in a star's velocity due to the gravitational tug of an orbiting planet. This method is particularly sensitive to large planets orbiting close in to their stars, and this planet was found to be closer to its star than Mercury is to the Sun, with a minimum mass about half that of Jupiter.

Twenty years later, in 2015, Kepler-452b was discovered during NASA's Kepler mission (Borucki et al. 2006), which found planets that pass in front of their host stars from our vantage point, blocking out a portion of the star's light, allowing the size of the planets to be calculated, since the transit depth measures the ratio of the areas of the planet and the star. Kepler-452b was one of the smallest planets found at the time, measuring in at 1.5 times the radius of the Earth, putting 20 years of progress in stark relief, as shown in the scale artist's concept in Figure 1. Several years later, we have identified systems of planets close in size to the Earth, such as the TRAPPIST-1 planets (Gillon et al. 2017), all seven of which are Earth-sized.

Kepler operated for more than nine years, and over that time, the scientific yield was extraordinary, having found close to 3000 planets on its own. Its successor, the *Transiting Exoplanet Survey Satellite* (*TESS*; Ricker et al. 2009), will likely dwarf this number in its search for planets outside of the solar system, also using the transiting technique. Covering an area of sky 400 times larger than that monitored by Kepler, *TESS* will survey 200,000 of the brightest stars in the solar neighborhood to search for transiting exoplanets. The stars *TESS* studies are 30–100 times brighter than those the Kepler mission and the *K*2 follow-up (Howell et al. 2014) surveyed, enabling far easier follow-up observations with both ground-based and space-based telescopes. *TESS* will, therefore, provide a critical link between NASA's Kepler mission and the *James Webb Space Telescope* (*JWST*; Gardner et al. 2006; Kalirai 2018), which will allow for characterization of the atmospheres of many nearby planets. In conjunction with dedicated efforts from programs on Extremely Large Telescopes on the ground, and next-generation space-based missions, such as the Large Ultraviolet (UV)/Optical/Infrared (IR) Surveyor (LUVOIR) and the Habitable Exoplanet Observatory (HabEx), the ability to characterize hundreds of transiting and directly imaged planets is within reach (Wang et al. 2018). The advances in instrumentation exhibited by these telescopes have the potential to revolutionize our understanding of all classes of extrasolar planets, particularly Earth-sized planets. The increased image resolution and wavelength coverage will allow us to probe the atmospheres of nearby potentially habitable worlds in search of water, molecular oxygen, ozone, carbon dioxide, methane, and other biosignatures—biologically produced global impacts to a planet's atmospheric and/or surface environment that can be remotely detected (Meadows 2005; Meadows et al. 2018; Schwieterman et al. 2018)—while, at the same time, making use of a wide range of adaptive optics techniques to hone in on directly imaging smaller, Earth-sized planets, as has already been done for larger, Jupiter-sized planets (see, e.g., Chauvin et al. 2004;

---
[*] Based on an invited talk at the 233rd meeting of the American Astronomical Society.

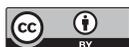 

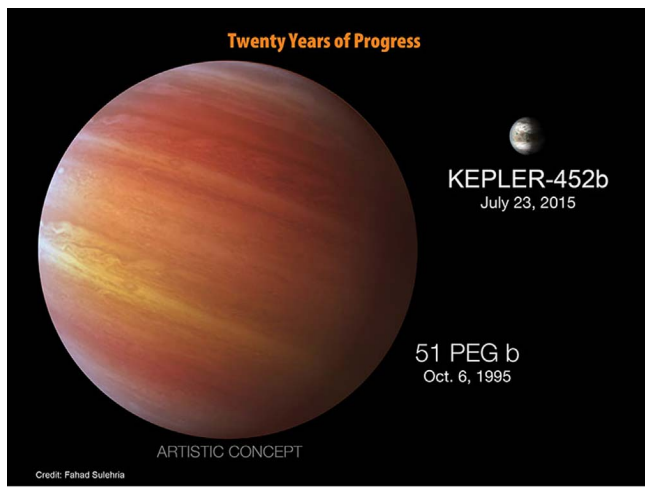

**Figure 1.** Artistic concept of the comparison between the first planet found orbiting another main-sequence star, 51 Pegasi b, and one of the smallest planets found at the time, Kepler-452b, measuring in at 1.5 times the radius of the Earth. Planets are drawn to scale. Credit: NASA Ames/W. Stenzel from https://exoplanets.nasa.gov/resources/271/twenty-years-of-progress/.

Marois et al. 2008, 2010). This broad scope of efforts is bringing us within visual range of the "holy grail" in the field of exoplanet research—the discovery of a habitable planet like the Earth that is capable of supporting and maintaining life over long timescales and perhaps even hosting life at the present time.

Among the most interesting recently discovered planets are: Proxima Centauri b (Anglada-Escudé et al. 2016), which is especially exciting because it orbits the star system nearest to us; the seven planets orbiting TRAPPIST-1 (Gillon et al. 2017), several of which orbit in the habitable zone (HZ) of their host star—the range of distances from its host star where an orbiting planet with an Earth-like atmosphere may be warm enough to sustain surface liquid water—(Kasting et al. 1993); and LHS 1140b, for which we have both mass and radius information (Dittmann et al. 2017), allowing confirmation of its rocky composition. While these planets certainly stoke the imagination regarding their potential to host life as we may (or may not) know it, the most revolutionary aspect of the young field of exoplanet research is the discovery that the solar system does not appear to be the standard model of solar systems in our Galaxy, but one of many possible configurations, including some that push at the boundaries of the traditional HZs, demanding that we consider the range of stellar and planetary parameters that are most influential for surface habitability.

In the coming years, hundreds more potentially habitable planets will be found by *TESS* and other observatories, leading to both an embarrassment of riches and a unique and challenging "problem": How do we identify the planets to look at more closely in our search for the next habitable planet where life exists, given limited telescope time on next-generation instrumentation? In this new era of large numbers of planet discoveries, there is a need for an approach that uses both observational data for newly discovered potentially habitable planets in combination with theoretical simulations using climate models. This combined approach will enable the identification of those planets that demonstrate the greatest likelihood of being habitable over a wide range of conditions and factors that are currently unexplored and for which no observational data currently exist.

Both observers and theorists alike who are on the search for life beyond the solar system look for planets that might be habitable—meaning they possess surface conditions conducive to the presence of liquid water (Seager 2013). Liquid water is one of the three fundamental requirements for life as we know it on the Earth, along with an energy source, and a suitable environment for the formation of complex organic molecules (Des Marais et al. 2008; Cockell et al. 2016). A terrestrial planet, by nature, has an energy source, along with the basic building blocks (in some form) that are needed for life, such as bioessential elements like sulfur, phosphorous, oxygen, nitrogen, carbon, and hydrogen (SPONCH; Hoehler 2007). What is not as common is liquid water, because it requires a particular climate capable of maintaining the requisite range of temperatures and pressures to maintain water in its liquid form. Life on Earth uses a diversity of metabolisms, and every one of them requires liquid water for chemical bonding and as a solvent for chemical reactions. Therefore, the presence of liquid water is the primary criterion directing the search for life beyond the Earth.

The long-term presence of surface liquid water on a planet is governed by the radiative and compositional properties of the host star, and the planet's individual properties, including its environment. The picture is rich and complex, as shown in Figure 2. A wide range of parameters and factors beyond orbital distance are of influence, few of which are constrained by observations. The majority of these parameters can only currently be explored using theoretical computer modeling, to aid in generating prioritized target lists of planets for follow-up observations. Over the past several years, researchers in the field of exoplanet climatology have explored many of these general areas of impact, using both observational data where available and climate models. This method has produced a deeper understanding of the likelihood of planets around certain spectral classes of stars to be habitable and is providing the most accurate assessments of the habitability of planets as they are discovered.

In the following sections, an overview of the use and its impact of computer models to understand the possible climates and potential habitability of extrasolar planets is provided. A general overview of the types of models used to explore planetary climate follows in Section 2. Section 3 describes the original use of climate models in predicting climate and weather patterns on the Earth. Section 4 outlines how these models have been applied to explore other planets in the solar system beyond the Earth. Sections 5 and 6 detail the eventual application of a hierarchy of climate models to the exploration of exoplanets, particularly those in the Earth-sized regime. Future work and conclusions follow in Sections 7 and 8.

## 2. Climate Models Used to Explore Planetary Climate

A climate model is a mathematical representation of a climate system based on its physical, biological, and chemical properties. The degree of detail with which a climate model addresses various climate forcings varies, yet all take into account the dominant initial contributors to a planet's climate: the incoming energy from the star and the outgoing radiation from the planet in response to that incoming energy.

The climates of other planets have been explored using a hierarchy of computer models, including radiative-convective (RC) climate models, energy balance models (EBMs)—some in concert with line-by-line radiative transfer (RT) models—and general circulation models, or global climate models (GCMs). These models span a wide range of complexity, from zero-dimensional (zero-D) to 3D, and each model has its advantages. Line-by-line RT models offer detailed treatments of atmospheric gas absorption. RC models include height as 1D and model the role of convection and vertical energy transfer, allowing the calculation of heat absorption in various atmospheric layers. Zero-D EBMs treat the entire planet as a single point in space, with an average surface temperature and outgoing energy (as shown in Figure 3), while 1D EBMs, which solve 1D energy balance equations, generate data that are zonally averaged over latitude (Figure 4). A chief advantage of EBMs is that they provide the ability to explore a wide parameter space without much computational expense, which is particularly important when addressing the climatic impacts of a range of factors like those depicted in Figure 2. The most complex models, with the highest resolution and the most sophisticated treatment of radiative transfer, atmospheric circulation, and ocean–atmosphere interactions, are 3D GCMs.

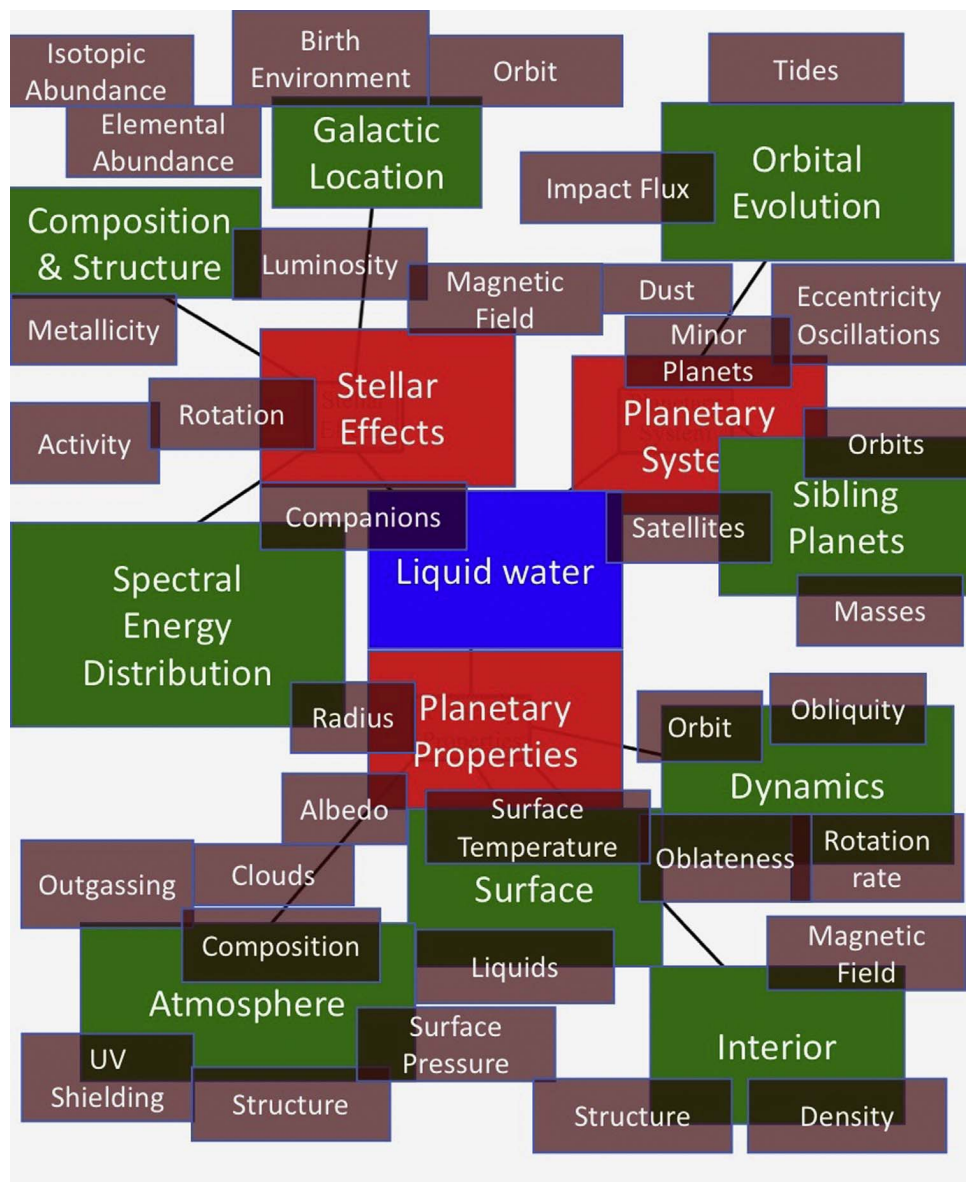

**Figure 2.** Schematic drawing of the multitude of factors and processes governing the long-term presence of liquid water on a planet's surface, based on Figure 1 from Meadows & Barnes (2018).

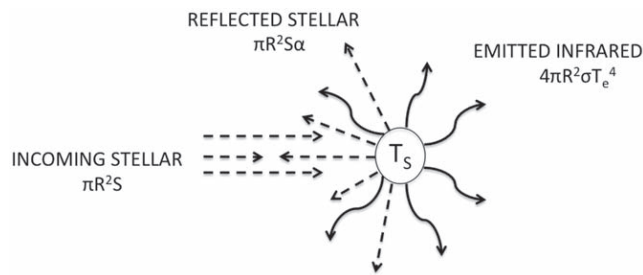

**Figure 3.** Schematic diagram of a zero-D energy balance model, based on Figure 3.1(a) in McGuffie & Henderson-Sellers (2005). Here the planet is treated as a single point in space, with a global mean effective temperature, $T_e$, and a surface temperature, $T_s$. If there are greenhouse gases in the atmosphere, then $T_s = T_e + \Delta T$. Here the emissivity, $\epsilon$, is assumed to be unity.

GCMs numerically solve a series of nonlinear differential equations describing atmospheric circulations through global fluid motions and transport in a dynamical core that characterizes resolved scale processes. In a 3D GCM, one starts with a sphere. On the surface of that sphere may be some fraction of ocean and land, some of which is covered by ice. And blanketing those planetary surfaces is an atmosphere. To represent the global climate system of such a planet in 3D then requires dividing the model planet into individual grid cells, with both horizontal (latitude/longitude) and vertical (height/pressure) components (Figure 5).[1] The model then calculates the average physical properties within each grid cell based on the physics and dynamics occurring within the cell itself and due to interactions with other cells. Such interactions include the radiative exchange between the incoming stellar energy, the atmosphere, and surface; atmospheric circulation, including winds; the exchange of heat between the atmosphere and the ocean; ocean circulation and its response to this heat exchange; and the response of ice and snow, clouds and water vapor, and soil and plants, through the exchange of water and heat between the surface and the atmosphere. These exchanges also include the cycling between the surface and atmosphere environments, particularly the cycling of gases, both volcanic and anthropogenic.

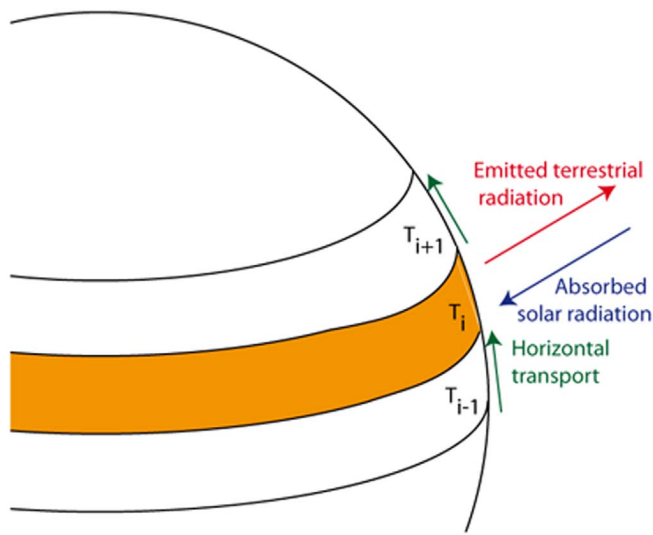

**Figure 4.** Schematic diagram of a 1D energy balance model based on Figure 3.1(b) in McGuffie & Henderson-Sellers (2005). The temperature is averaged over bands of latitude, based on the balance between absorbed solar and emitted terrestrial radiation, with horizontal heat transport from areas of energy surplus at the tropics to polar latitudes, where there is an energy deficit.

Not all GCMs include this full suite of processes. For simplicity, some modeling efforts assume an aqua planet (no land), for example. However, the range of physical processes described above provides testimony to the degree of detail climate models can attain in the attempt to accurately depict the global climate system of a planet, which has historically been the Earth orbiting the Sun. Efforts to apply climate models to other planets have directly benefited from early work using these models to understand and interpret climate and weather patterns on the Earth.

### 3. Using Climate Models to Explore the Earth

Numerical models like those described in Section 2 have been employed to identify the major factors and processes shaping Earth's climate. Central to these studies were efforts to understand the manner in which global energy balance is achieved through a combination of reflected, absorbed, and/or emitted shortwave (incoming stellar) and longwave (outgoing thermal) radiation. Such model applications began in the 1960s. In seminal works, Budyko (1969) and Sellers (1969) presented the first early EBMs. Their models of the annual temperature distribution on Earth were based on the principle that the energy absorbed is balanced by the energy emitted by the planet, with any imbalance in the system resulting in a change in the surface temperature. They found the latitudinally averaged temperature distribution (Figure 6) to be affected by climate feedbacks, such as ice-albedo feedback—a positive feedback caused by the contrast between the reflectivity of ice cover and ocean. Within a similar timeframe, GCMs were employed to understand and also to predict the future climate on the Earth (see, e.g., Manabe et al. 1965; Smagorinsky et al. 1965; Holloway & Manabe 1971; Manabe & Wetherald 1975) and are currently being used to forecast the global impact of anthropogenic carbon dioxide emission-induced climate change into the 2100s, as shown in Figure 7.

Later, climate models, from 1D to 3D, began to apply knowledge of ice-albedo and other climate feedbacks to problems related to Earth's early climate, including the faint young Sun paradox (Kasting et al. 1984; Haqq-Misra et al. 2008; Charnay et al. 2013; Wolf & Toon 2013) and the question of how Earth exited the Snowball episodes hypothesized to have occurred ~715 and ~630 Ma (Kirschvink 1992), given constraints on atmospheric $CO_2$ levels at the time (Pierrehumbert et al. 2011). One of the proposed mechanisms involved a possible "Mudball" rather than Snowball state on the Neoproterozoic Earth, with ice and snow surfaces made darker and more absorbing (thus easier to thaw) by the presence of continental dust (Abbot & Pierrehumbert 2010). Additional work explored alternative climate scenarios on the Earth that might have explained the persistence of photosynthetic life throughout these global-scale glaciations (Knoll 1992), including a "Jormungand" state, which is characterized by a narrow belt of open water in the tropics and is stable in model simulations as a consequence of the lower albedo contrast between bare sea ice and the ocean (Abbot et al. 2011). If such a state could be possible on the Earth, planets in other systems could exist in similar "waterbelt" states (see, e.g., Wolf et al. 2017). And closer to home, climate models could be used to explore other planets within the solar system, including Mars, host to another planetary climate mystery, and Venus, whose deep atmosphere contained previously uncharted regions waiting to be mapped.

### 4. Solar System Studies

The application of climate models to other planets in the solar system beyond the Earth began in the late 1990s. A key question of interest during this time was how Mars could have been warm enough for liquid water to flow on its surface, as abundant physical evidence indicates (Fassett & Head 2008; Hynek et al. 2010; Grotzinger et al. 2015), given the 20%–25% lower luminosity of the Sun (Wordsworth et al. 2017). Initial model simulations proposed backscattering of $CO_2$ ice grains, depending on particle size and optical depth, as a potential mechanism for generating above-freezing surface temperatures on Mars (Forget & Pierrehumbert 1997). Subsequent modeling efforts found this effect to have been previously overestimated (Colaprete & Toon 2003; Forget et al. 2013; Kitzmann 2016).

---
[1] https://celebrating200years.noaa.gov/breakthroughs/climate_model/modeling_schematic.html

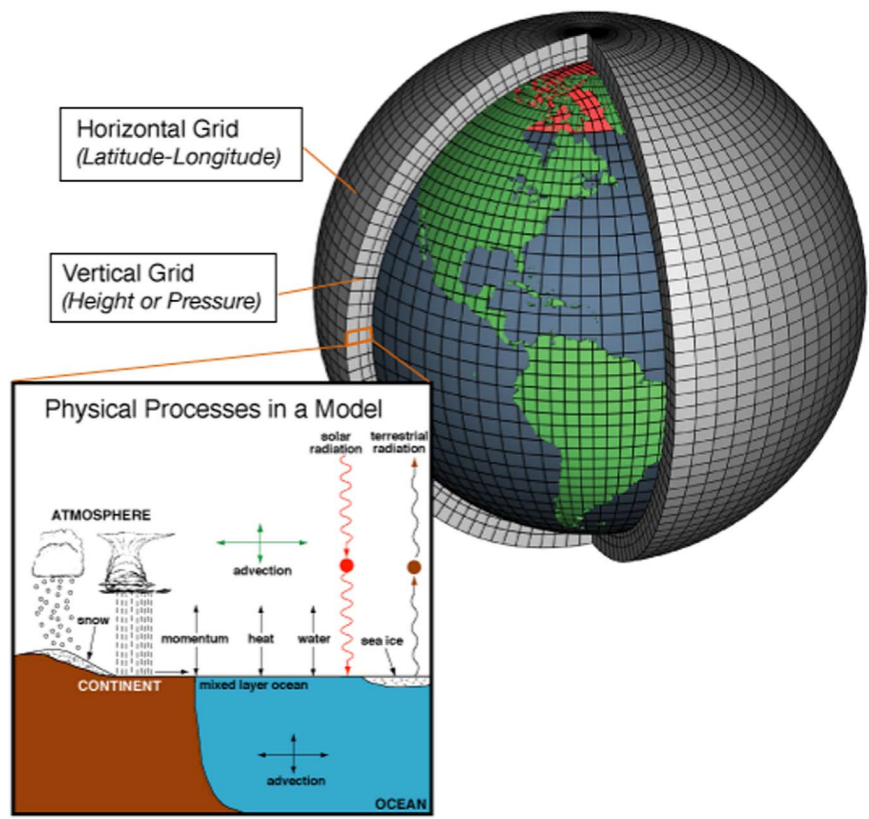

**Figure 5.** Schematic diagram of a 3D general circulation model. The planet is divided into horizontal and vertical grids, and individual properties of the climate system are evaluated within each grid cell.

More recent work suggests a new mechanism—$CO_2$–$H_2$ and $CO_2$–$CH_4$ collision-induced absorption—to be the most plausible explanation, as both methane and hydrogen, generated as a result of aqueous crustal alteration, could have combined with $CO_2$ outgassed from volcanoes to form temporal atmospheres during the period when liquid water flowed on the surface of Mars, approximately one billion years ago (Wordsworth et al. 2017).

GCMs have been applied to planets with massive, slowly rotating atmospheres, like that of Venus (Rossow 1983), as well as to planetary bodies with much colder extreme environments, like those of Titan (Friedson et al. 2009; Charnay et al. 2015a; Larson et al. 2015) and Pluto (Forget et al. 2017). New models, such as the "Laboratoire de Météorologie Dynamique" (LMD) Generic GCM (Hourdin et al. 2006), and the Resolving Orbital and Climate Keys of Earth and Extraterrestrial Environments with Dynamics (ROCKE-3D) GCM (Way et al. 2017) were created with the express purpose of studying non-Earth-like planets, with varied radii, surface gravities, surface pressures, and atmospheric compositions and resulting chemistries. These models took into account specific characteristics of solar system planets beyond the Earth, such as Venus' specific topography and diurnal cycle, thereby permitting an understanding of its dynamics, including the detailed structure of its wind patterns and the superrotation of its atmosphere (e.g., Lebonnois et al. 2010, 2018; Ando et al. 2016). This specialized treatment of the environments of other planets in the solar system paved the way for the adjustment of Earth-focused climate models to incorporate the characteristics of different stellar environments as well. With these modifications, modeling efforts could journey much farther afield to explore and assess the potential climates of planets orbiting stars other than the Sun.

## 5. Exoplanets: General Parameter Studies

The HZ was defined and quantified as a function of stellar mass in the early 1990s by Kasting et al. (1993). The primary step in categorizing a planet as "potentially habitable"—the identification of a planet orbiting in the HZ of its host star—is long understood to be an important step in the process of assessing a planet's habitability potential. However, the Snowball Earth episodes are a prime example of a planet that likely experienced a significant climatic event that would certainly have impacted the long-term presence of surface liquid water, emphasizing the wide range of factors beyond orbital distance (as illustrated in Figure 2) that are important to the discussion of planetary habitability.

Modeling efforts have addressed the impact on habitability of many of these factors, with a focus on how, given their effects on Earth's climate, these climatic impacts might change for different stellar and planetary environments. In particular, the effects of extreme values of orbital elements on the climate of exoplanets have been explored, including obliquity (Williams & Kasting 1997; Williams & Pollard 2003; Spiegel et al. 2009; Ferreira et al. 2014), which increases seasonality and globally averaged surface temperatures (due to the more direct angle at which the Sun's rays hit the planet), obliquity oscillations (Armstrong et al. 2014), which were shown in simulations to prevent thick ice sheets from forming on either hemisphere throughout an orbit, increasing habitable surface area on planets exhibiting such behavior, and eccentricity (Williams & Pollard 2002; Dressing et al. 2010; Bolmont et al. 2016), which increases annually averaged flux received by a planet. The role of land versus ocean surfaces has also been explored on planets orbiting Sun-like stars (Abe et al. 2011; Foley 2015), yielding a deepening understanding of how the climate of land planets may differ markedly from that of planets covered predominantly by water.

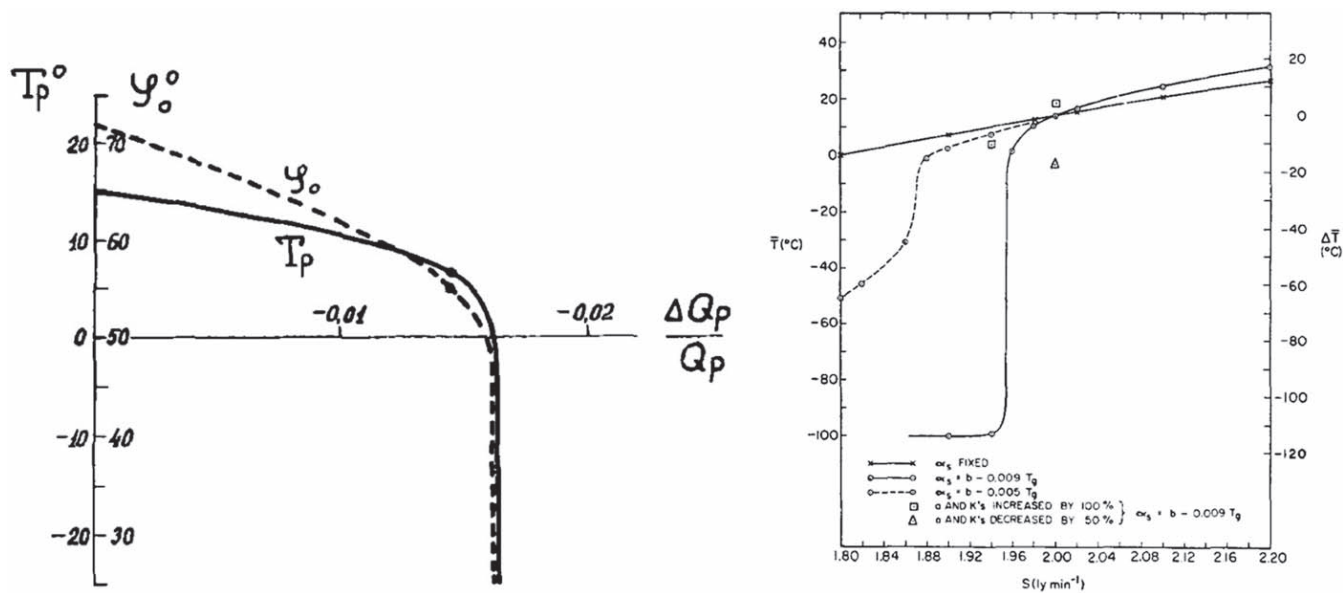

**Figure 6.** Left: Figure 5 from Budyko (1969) showing the dependence of Earth's temperature and ice line latitude on incoming solar radiation. Copyright 1969 M.I. Budyko. Published by Taylor and Francis Group LLC. Used with permission. Right: Figure 6 from Sellers (1969), showing the mean global temperature at sea level on the Earth, as a function of the solar constant, for different responses in planetary albedo, and different meridional exchange coefficient and eddy diffusivity values. Copyright 1969 American Meteorological Society. Used with permission.

Climate simulations of exoplanets have put a great deal of focus on quantifying the effects on climate of the unique stellar environments of orbiting planets. For example, M-dwarf planets, given the close proximity of their host stars' HZs, may be captured into resonances in spin–orbit periods, including the extreme case of synchronous rotation (Dole 1964). The atmospheric effects of this 1:1 spin–orbit resonance have been studied using GCMs, unearthing a number of interesting results that counter earlier concerns about the frigid temperatures and the possibility of atmospheric collapse on the night sides of these planets. Work on this topic has proposed fairly reasonable atmospheric requirements (a 100 mb $CO_2/H_2O$ atmosphere) for sufficient heat transport between the day and night sides to avoid such a fate (Haberle et al. 1996; Joshi et al. 1997; Wordsworth 2015), as well as a stabilizing cloud feedback (Yang et al. 2013, 2014) that could buffer slowly rotating planets with oceans from entering into "runaway greenhouse" states (see, e.g., Ingersoll 1969) too close in to the HZ. Indeed, the effect on climate of planetary rotation rate has garnered a great deal of attention (Edson et al. 2011; Kite et al. 2011; Showman et al. 2013; Hu & Yang 2014; Wang et al. 2014a, 2014b), particularly as it pertains to the synchronous case, as its effect on atmospheric circulation will influence the climatic impacts of other previously explored factors, such as land fraction and distribution (Edson et al. 2012; Leconte et al. 2013). And the climatic effects of ocean dynamics have also been explored for slowly rotating M-dwarf planets (see, e.g., Del Genio et al. 2019; Yang et al. 2019).

The effect on planetary climate of the interaction between the spectral energy distribution of host stars, and their planets' atmospheres and surfaces has been an area of recent study (Shields et al. 2013, 2014; von Paris et al. 2013; Godolt et al. 2015; Wolf et al. 2017). The lower albedo of ice and snow in the near-IR has led to a quantified understanding of the greater potential climate stability of planets orbiting stars with more near-IR output (Shields et al. 2013, 2014). And a newly studied surface type—sodium chloride dihydrate, or "hydrohalite" ($NaCl \cdot 2H_2O$), which can crystallize in bare sea ice and is highly reflective in the near-IR—has been found to generate colder global mean surface temperatures on HZ M-dwarf (and G-dwarf) planets when albedo parameterizations for its formation are included in climate simulations (Shields & Carns 2018). Computer models that include photochemistry have also been useful in providing an understanding of the robustness of an ozone shield on planets subjected to single flares from M-dwarf host stars (Segura et al. 2003, 2010) and of the more detrimental effect on ozone column depth and atmospheric retention in the presence of repeated flare events and in absence of a planetary magnetic field (Tilley et al. 2019).

These previous efforts to quantify the effects on climate and habitability of a range of parameters known to influence the long-term presence of liquid water on a planet's surface have been crucial to a greater understanding of what is truly necessary for surface habitability on an Earth-like planet. General studies of parameters that display significant departures from Earth-like behavior when applied to different stellar and planetary environments have allowed for subsequent applications of climate models to comprehensive assessments of the potential habitability of recently discovered planets.

## 6. Exoplanets: Targeted Planet Studies

The use of climate models to quantify the climatic impact of many different variables has been pivotal to the facility of these models in tackling new hurdles in the field—specifically, their application to actual, observed planetary systems. Over the past decade, the use of 3D GCMs, in particular, has expanded to apply this quantified understanding of climatic effects to recently discovered exoplanets, producing assessments of their possible climates and potential habitability that take the specific planetary system into account, including incorporating observational data where available.

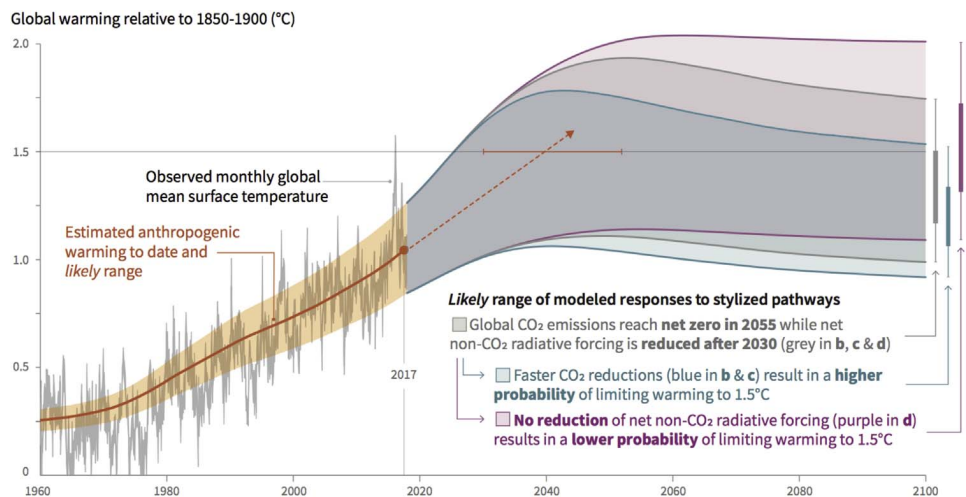

**Figure 7.** Global warming relative to 1850–1900, from the Intergovernmental Panel on Climate Change (IPCC) 2018 summary for policymakers (Masson-Delmotte et al. 2018).

One of the first of such efforts was the identification of scenarios that would allow for open water on the planet Gliese 581g, including an "Eyeball Earth" state, with open water at the substellar point of an otherwise frozen synchronously rotating planet (Pierrehumbert 2011), as shown in Figure 8. Other planets in the same system have been explored, and their habitability has been quantified as a function of atmospheric composition and rotation rate (Selsis et al. 2007; Wordsworth et al. 2010).

Recent works have used GCMs to identify a range of possible climate scenarios as a function of similar factors for Proxima Centauri b (Turbet et al. 2016; Boutle et al. 2017; Del Genio et al. 2019), as shown, for example, in Figure 9. Similar work has been done for the TRAPPIST-1 planets (Wolf 2017, 2018; Turbet et al. 2018). The effect on surface habitability of the gravitational interactions that can occur in multiple-planet systems has been explored using constraints from *n*-body models as inputs to GCMs for the first time, with a study of the effect of orbital configuration on the requirements for open water on the potentially habitable planet Kepler-62f, for a range of high- and low-$CO_2$ cases and for different rotation periods (Figure 10; Shields et al. 2016). And in parallel, GCMs have been used to explore planets with likely climates that are very different from the Earth, including planets in the Jupiter-mass regime orbiting close in to their stars (so-called "Hot Jupiters"; see, e.g., Showman et al. 2008a, 2008b, 2009; Menou & Rauscher 2009; Fortney et al. 2010; Rauscher & Menou 2010; Amundsen et al. 2014, 2016; Mayne et al. 2014; Kataria et al. 2015; Jiménez-Torres 2016; Roman & Rauscher 2017, 2019), the warm sub-Neptune GJ1214b (Charnay et al. 2015b, 2015c), and the potential Venus analog Kepler-1649b (Kane et al. 2018). All told, these targeted planet case studies have brought to light the wide variety of conditions that might exist outside of the solar system. This theoretical work has also provided knowledge about the stellar and planetary environments most conducive to planetary surface habitability, which is key to directing the crucial next steps in the field.

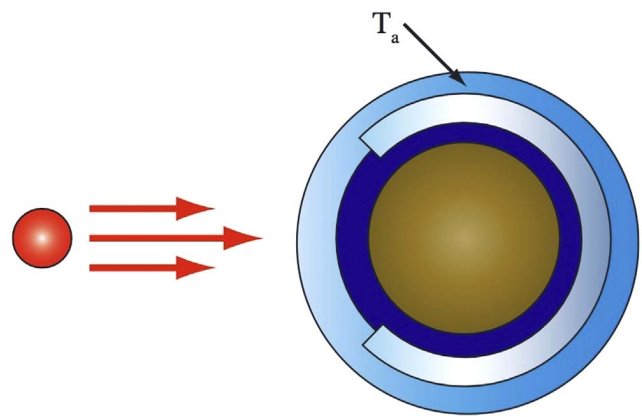

**Figure 8.** Left: the Eyeball Earth climate state, where there is open water at the substellar point and a frozen surface across the rest of the planet (Perrehumbert 2011). In this schematic, $T_a$ refers to the near-surface temperature of the atmosphere.

### 7. The Future of the Field

As exoplanet discoveries increase in the coming years as a result of *TESS* and ground-based efforts, the field will continue to widen to include detailed characterization of these worlds. As a result, theoretical modeling will become even more critical to the search for habitable planets and life beyond the solar system. Combining the use of observational data for newly discovered planets with theoretical computer modeling will produce accurate characterizations of the most scientifically interesting prospects for habitable planets. These case studies are helping to build a prioritized target list of planets to follow up on with the next generation of space-based missions, such as *JWST*, LUVOIR, and HabEx, which will attempt to measure the atmospheric composition of Earth-sized planets, in hopes of detecting the presence of life. While life may exist in oceans sequestered beneath thick ice sheets on other planets or moons, it is those planets where life has the potential to thrive on the surface that are of prime interest to the observational exoplanet community, as such life would possess the greatest capability of making its presence known in the atmosphere, on the surface, or in other ways that might be measured remotely by space telescopes in search of biosignatures. Future spectroscopic missions will likely have a long list of planets to choose from, and theoretical modeling efforts will be critical to making those choices.

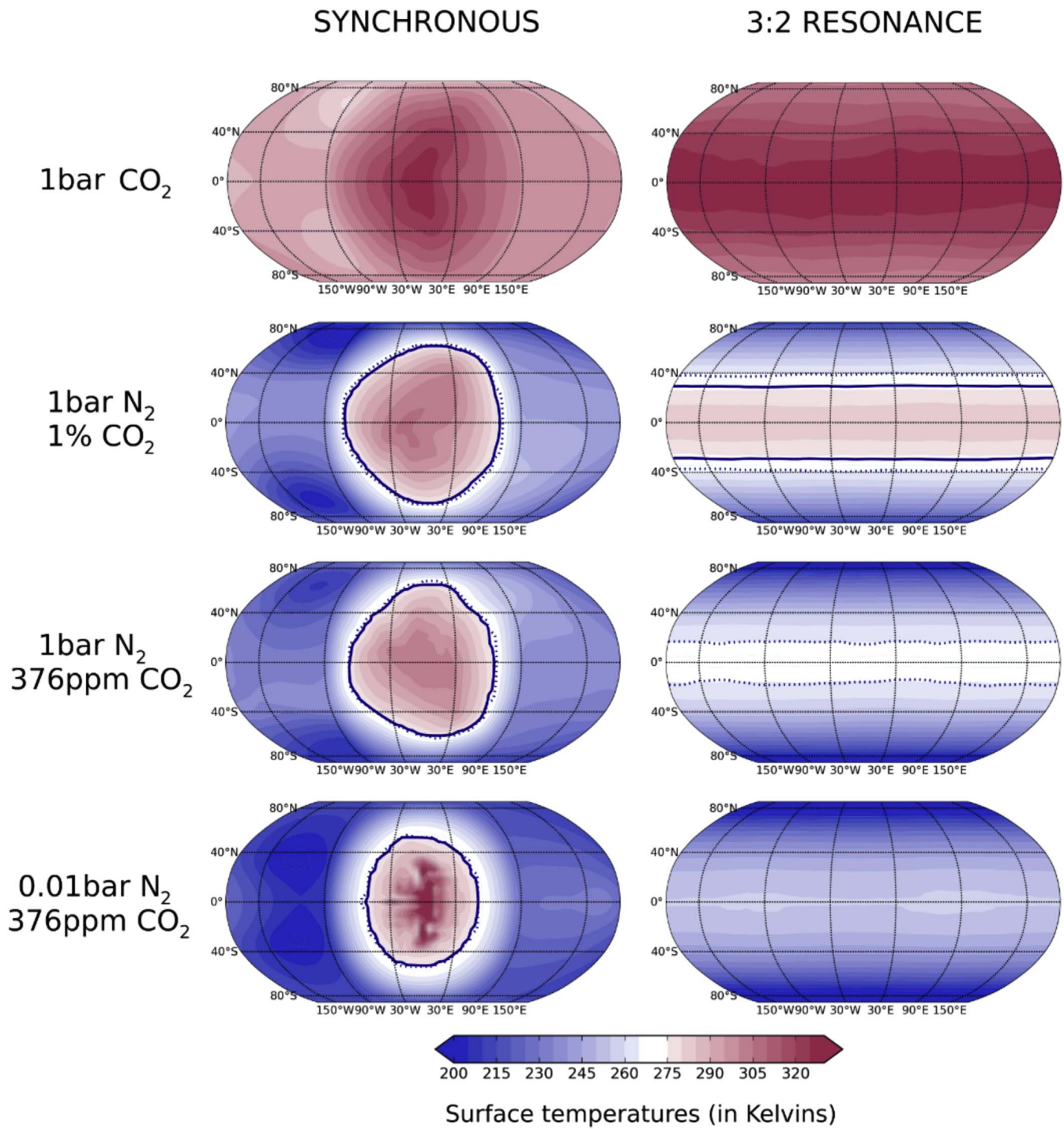

**Figure 9.** Left: Figure 6 from Turbet et al. (2016) showing a variety of potential climates for an ocean-covered Proxima Centauri b, assuming a range of atmospheric compositions of $N_2$ and $CO_2$ for a synchronously rotating case (1:1 spin–orbit resonance) and one with a 3:2 spin–orbit resonance. Copyright 2016 ESO. Reproduced with permission from ESO.

The gap between the theoretical modeling and observational exoplanet communities has narrowed appreciably due to complementary efforts. Climate models are now being used to simulate suites of phase curves—diagrams displaying the brightness of a planet as a function of its phase angle—and other observables, like those for TRAPPIST-1e shown in Figure 11. This relatively recent application of GCMs will help astronomers determine the impact on observational signatures of different climatic variables, such as atmospheric and surface composition, and usher in the ability to identify specific planetary conditions as a function of interpreted observational measurements. Hot Jupiters are currently the only planets with observational measurements that are detailed enough to compare at the most basic level with 3D modeled observational signatures (see, e.g., Kreidberg et al. 2018). For terrestrial planets, this task is much more challenging, and likely an effort for which satisfying observations will remain a future goal. In the meantime, simulating these observational signatures will offer the chance to produce the most interdisciplinary and comprehensive portraits of potentially habitable worlds to date. Far surpassing traditional discovery papers, these new characterizations will incorporate observational data on newly discovered planets alongside simulations of possible climate regimes as a function of unmeasured planetary properties, as well as the computed observational signatures of those properties. The end result of those efforts stands to be an understanding of the full extent of scenarios possible and necessary, both theoretically and observationally, to produce and sustain a habitable planet.

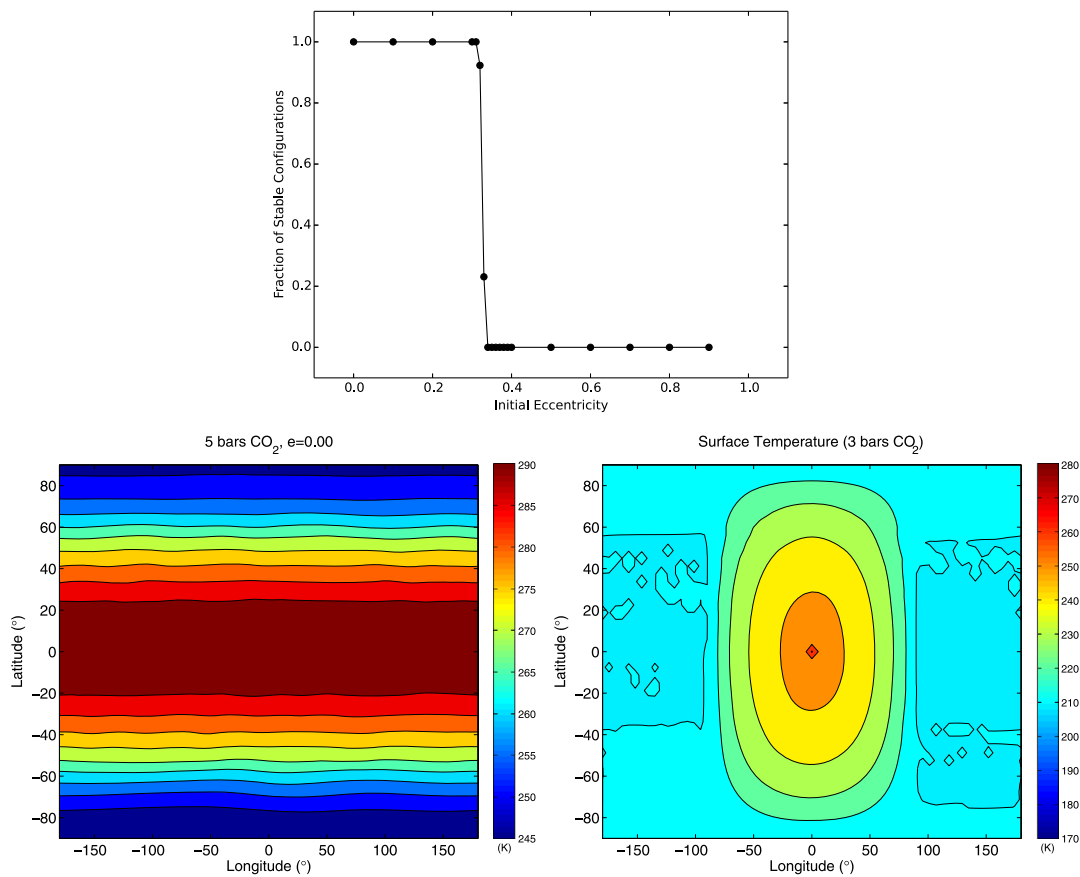

**Figure 10.** Top: Figure 2 from Shields et al. (2016) showing the fraction of stable configurations possible for Kepler-62f as a function of initial eccentricity, using the Hierarchical N-Body (HNBody) package (Rauch & Hamilton 2002). Bottom: Figures 9 (left) and 10(c) (right) from Shields et al. (2016), showing surface temperature as a function of latitude, assuming the lowest eccentricity constraint from the n-body model; five and three bars of $CO_2$, respectively; and 24 hr and synchronous rotation rates, respectively. Copyright 2016 Mary Ann Liebert Inc. Reproduced with permission from Mary Ann Liebert, Inc., New Rochelle, NY.

## 8. Conclusions

The relatively new field of exoplanet climatology—a field dedicated to applying the principles that govern climate on the Earth to the exploration of the possible climates and potential habitability of extrasolar planets—has a long history. Starting with the first early modeling of important climatic feedbacks on the Earth, climate models have developed and evolved to predict its climate and weather patterns into the 2100s, to explain atmospheric superrotation and other phenomena on planets elsewhere in the solar system, and finally, to assess the potential habitability of newly discovered exoplanets. The three primary uses of climate models across the hierarchical plane to explore exoplanet climates and habitability have taken the field of exoplanet climatology from purely theoretical to both applicative and essential to the characterization of new worlds. Modeling the effect on climate of a range of parameters unconstrained by observations has helped to identify the likelihood of planets orbiting certain spectral classes of stars to be habitable. Applying this general knowledge of the influence on habitability of unique stellar and planetary environments to actually observed planetary systems will facilitate the real-time assessment of a planet's habitability potential as it is discovered, laying the groundwork for discovery papers that move beyond observational data alone and to combine this information with quantitative determinations of the possible climates of those planets. Finally, computing the impacts of factors, such as atmospheric and surface composition, and rotation rate on observational measurements will allow astronomers to compare the data collected for newly discovered planets with modeled observations, disentangling competing effects to accurately determine the precise conditions on a planet. Combined with those conditions we know to be essential for habitability and life, this information will hopefully one day help one planet emerge above the rest as the planet that has it all—water, organics, energy, and indeed, life.

This material is based upon work supported by NASA under grant number NNH16ZDA001N, which is part of the "Habitable Worlds" program; by the National Science

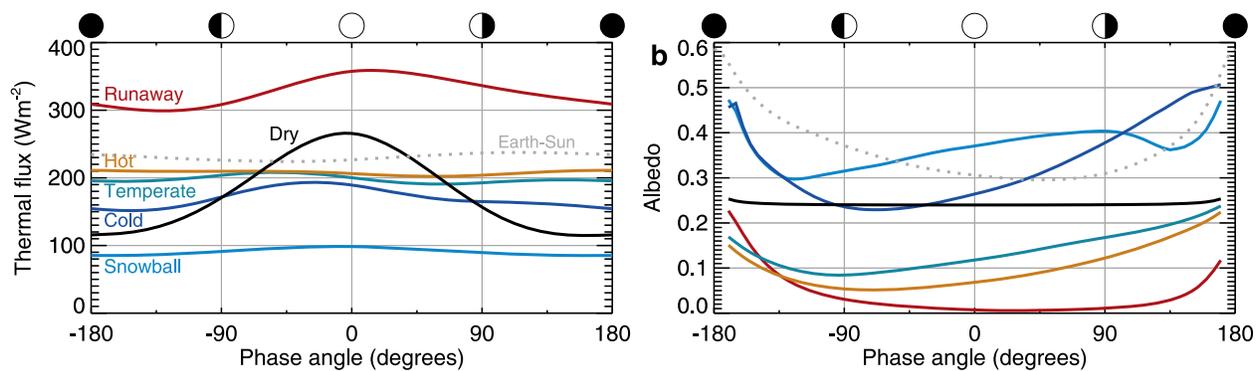

**Figure 11.** Thermal flux (left) and planetary albedo as a function of phase angle, highlighting different possible climate regimes on TRAPPIST-1 planets, updated from Figure 5 from Wolf (2017) to include a dry planet case with a 1 bar $N_2$ atmosphere and no $H_2O$, and the Earth–Sun case for comparison. Credit: Wolf (2017).

Foundation under Award No. 1753373; and by a Clare Boothe Luce Professorship. I would like to acknowledge high-performance computing support from Yellowstone (ark:/85065/d7wd3xhc) and Cheyenne (doi:10.5065/D6RX99HX) provided by NCAR's Computational and Information Systems Laboratory, sponsored by the National Science Foundation. This work was performed as part of the NASA Astrobiology Institute's Virtual Planetary Laboratory under Cooperative Agreement No. NNA13AA93A. I thank Michael Strauss, Chick Woodward, James Lowenthal, and the American Astronomical Society for the invitation to present a plenary talk at the 233rd Winter Meeting in Seattle, Washington that inspired this review, and Ethan Vishniac and the AAS journals for the invitation to write this review. Many thanks to Raymond Pierrehumbert and Julianne Dalcanton for helpful correspondence and Eric Wolf for reading an earlier version of this manuscript and providing essential feedback.

### ORCID iDs

Aomawa L. Shields  https://orcid.org/0000-0002-7086-9516

### References

Abbot, D. S., & Pierrehumbert, R. T. 2010, JGRD, 115, D03104
Abbot, D. S. G., Voigt, A., & Koll, D. 2011, JGRD, 116, D18103
Abe, Y., Abe-Ouchi, A., Sleep, N. H., & Zahnle, K. J. 2011, AsBio, 11, 443
Amundsen, D. S., Baraffe, I., Tremblin, P., et al. 2014, A&A, 564, A59
Amundsen, D. S., Mayne, N. J., Baraffe, I., et al. 2016, A&A, 595, A36
Ando, H., Sugimoto, N., Takagi, M., et al. 2016, NatCo, 7, 10398
Anglada-Escudé, G., Amado, P. J., Barnes, J., et al. 2016, Natur, 536, 437
Armstrong, J. C., Barnes, R., Domagal-Goldman, S., et al. 2014, AsBio, 14, 277
Bolmont, E., Libert, A.-S., Leconte, J., & Selsis, F. 2016, A&A, 591, A106
Borucki, W. J., Koch, D., Basri, G., et al. 2006, ISSIR, 6, 207
Boutle, I. A., Mayne, N. J., Drummond, B., et al. 2017, A&A, 601, A120
Budyko, M. I. 1969, Tell, 21, 611
Charnay, B., Barth, E., Rafkin, S., et al. 2015a, NatGe, 8, 362
Charnay, B., Forget, F., Wordsworth, R., et al. 2013, JGRD, 118, 10414
Charnay, B., Meadows, V., & Leconte, J. 2015b, ApJ, 813, 15
Charnay, B., Meadows, V., Misra, A., Leconte, J., & Arney, G. 2015c, ApJL, 813, L1
Chauvin, G., Lagrange, A.-M., Dumas, C., et al. 2004, A&A, 425, L29
Cockell, C. S., Bush, T., Bryce, C., et al. 2016, AsBio, 16, 89
Colaprete, A., & Toon, O. B. 2003, JGRE, 108, 5025
Del Genio, A. D., Way, M. J., Amundsen, D. S., et al. 2019, AsBio, 19, 99
Des Marais, D. J., Nuth, J. A., III, Allamandola, L. J., et al. 2008, AsBio, 8, 715
Dittmann, J. A., Irwin, J. M., Charbonneau, D., et al. 2017, Natur, 544, 333
Dole, S. H. 1964, Habitable Planets for Man (New York: Blaisdell)
Dressing, C. D., Spiegel, D. S., Scharf, C. A., Menou, K., & Raymond, S. N. 2010, ApJ, 721, 1295
Edson, A., Lee, S., Bannon, P., Kasting, J. F., & Pollard, D. 2011, Icar, 212, 1
Edson, A. R., Kasting, J. F., Pollard, D., Lee, S., & Bannon, P. R. 2012, AsBio, 12, 562
Fassett, C. I., & Head, J. W. 2008, Icar, 198, 37
Ferreira, D., Marshall, J., O'Gorman, P. A., & Seager, S. 2014, Icar, 243, 236
Foley, B. J. 2015, ApJ, 812, 36
Forget, F., Bertrand, T., Vangvichith, M., et al. 2017, Icar, 287, 54
Forget, F., & Pierrehumbert, R. T. 1997, Sci, 278, 1273
Forget, F., Wordsworth, R., Millour, E., et al. 2013, Icar, 222, 81
Fortney, J. J., Shabram, M., Showman, A. P., et al. 2010, ApJ, 709, 1396
Friedson, A. J., West, R. A., Wilson, E. H., Oyafuso, F., & Orton, G. S. 2009, P&SS, 57, 1931
Gardner, J. P., Mather, J. C., Clampin, M., et al. 2006, SSRv, 123, 485
Gillon, M., Triaud, A. H. M. J., Demory, B.-O., et al. 2017, Natur, 542, 456
Godolt, M., Grenfell, J. L., Hamann-Reinus, A., et al. 2015, P&SS, 111, 62
Grotzinger, J. P., Gupta, S., Malin, M. C., et al. 2015, Sci, 350, 7575
Haberle, R. M., McKay, C. P., Tyler, D., & Reynolds, R. T. 1996, in Proc First Int. Conf. on Circumstellar Habitable Zones, ed. L. R. Doyle (Menlo Park, CA: Travis House Publications), 29
Haqq-Misra, J. D., Domagal-Goldman, S. D., Kasting, P. J., & Kasting, J. F. 2008, AsBio, 8, 1127
Hoehler, T. M. 2007, AsBio, 7, 824
Holloway, J. L., & Manabe, S. 1971, MWRv, 99, 335
Hourdin, F., Musat, I., Bony, S., et al. 2006, ClDy, 27, 787
Howell, S. B., Sobeck, C., Haas, M., et al. 2014, PASP, 126, 398
Hu, Y., & Yang, J. 2014, PNAS, 111, 629
Hynek, B. M., Beach, M., & Hoke, M. R. T. 2010, JGRE, 115, E09008
Ingersoll, A. P. 1969, JAtS, 26, 1191
Jiménez-Torres, J. J. 2016, RMxAA, 52, 69
Joshi, M. M., Haberle, R. M., & Reynolds, R. T. 1997, Icar, 129, 450
Kalirai, J. 2018, ConPh, 59, 251
Kane, S. R., Ceja, A. Y., Way, M. J., & Quintana, E. V. 2018, ApJ, 869, 46
Kasting, J. F., Pollack, J. B., & Crisp, D. 1984, JAtC, 1, 403
Kasting, J. F., Whitmire, D. P., & Reynolds, R. T. 1993, Icar, 101, 108
Kataria, T., Showman, A. P., Fortney, J. J., et al. 2015, ApJ, 801, 86
Kirschvink, J. 1992, in The Proterozoic Biosphere: A Multidisciplinary Study, ed. J. Schopf (Cambridge: Cambridge Univ. Press), 51, https://authors.library.caltech.edu/36446/
Kite, E. S., Gaidos, E., & Manga, M. 2011, ApJ, 743, 41
Kitzmann, D. 2016, ApJL, 817, L18
Knoll, A. 1992, Sci, 256, 622
Kreidberg, L., Line, M. R., Parmentier, V., et al. 2018, AJ, 156, 17
Larson, E. J. L., Toon, O. B., West, R. A., & Friedson, A. J. 2015, Icar, 254, 122
Lebonnois, S., Hourdin, F., Eymet, V., et al. 2010, JGRE, 115, E06006
Lebonnois, S., Schubert, G., Forget, F., & Spiga, A. 2018, Icar, 314, 149
Leconte, J., Forget, F., Charnay, B., et al. 2013, A&A, 554, A69
Manabe, S., Smagorinsky, J., & Strickler, R. F. 1965, MWRv, 93, 769
Manabe, S., & Wetherald, R. T. 1975, JAtS, 32, 3
Marcy, G. W., Butler, R. P., Williams, E., et al. 1997, ApJ, 481, 926
Marois, C., Macintosh, B., Barman, T., et al. 2008, Sci, 322, 1348
Marois, C., Zuckerman, B., Konopacky, Q. M., Macintosh, B., & Barman, T. 2010, Natur, 468, 1080
Masson-Delmotte, V., Zhai, P., Pörtner, H.-O., et al. 2018, Global Warming of 1°5C. An IPCC Special Report on the Impacts of Global Warming of 1°5C above Pre-industrial Levels and Related Global Greenhouse Gas Emission

Pathways, in the Context of Strengthening the Global Response to the Threat of Climate Change, Sustainable Development, and Efforts to Eradicate Poverty, (Geneva: World Meteorological Org.)
Mayne, N. J., Baraffe, I., Acreman, D. M., et al. 2014, A&A, 561, A1
Mayor, M., & Queloz, D. 1995, Natur, 378, 355
McGuffie, K., & Henderson-Sellers, A. 2005, A History of and Introduction to Climate Models (New York: Wiley)
Meadows, V. S. 2005, in Proc. IAU Colloq. 200, Direct Imaging of Exoplanets: Science & Techniques, ed. C. Aime & F. Vakili (Cambridge: Cambridge Univ. Press), 25
Meadows, V. S., Arney, G. N., Schwieterman, E. W., et al. 2018, AsBio, 18, 133
Meadows, V. S., & Barnes, R. K. 2018, in Handbook of Exoplanets, ed. H. J. Deeg & J. A. Belmonte (New York: Springer Nature), 2771
Menou, K., & Rauscher, E. 2009, ApJ, 700, 887
Pierrehumbert, R. T. 2011, ApJL, 726, 8
Pierrehumbert, R. T., Abbot, D. S., Voigt, A., & Koll, D. 2011, AREPS, 39, 417
Rauch, K. P., & Hamilton, D. P. 2002, BAAS, 34, 938
Rauscher, E., & Menou, K. 2010, ApJ, 714, 1334
Ricker, G. R., Latham, D. W., Vanderspek, R. K., et al. 2009, AAS Meeting, 213, 403.01
Roman, M., & Rauscher, E. 2017, ApJ, 850, 17
Roman, M., & Rauscher, E. 2019, ApJ, 872, 1
Rossow, W. B. 1983, JAtS, 40, 273
Schwieterman, E. W., Kiang, N. Y., Parenteau, M. N., et al. 2018, AsBio, 18, 663
Seager, S. 2013, Sci, 340, 577
Segura, A., Krelove, K., Kasting, J. F., et al. 2003, AsBio, 3, 689
Segura, A., Walkowicz, L. M., Meadows, V., Kasting, J., & Hawley, S. 2010, AsBio, 10, 751
Sellers, W. D. 1969, JApMe, 8, 392
Selsis, F., Kasting, J. F., Levrard, B., et al. 2007, A&A, 476, 1373
Shields, A. L., Barnes, R., Agol, E., et al. 2016, AsBio, 16, 443
Shields, A. L., Bitz, C. M., Meadows, V. S., Joshi, M. M., & Robinson, T. D. 2014, ApJL, 785, L9
Shields, A. L., & Carns, R. C. 2018, ApJ, 867, 11
Shields, A. L., Meadows, V. S., Bitz, C. M., et al. 2013, AsBio, 13, 715
Showman, A. P., Cooper, C. S., Fortney, J. J., & Marley, M. S. 2008a, ApJ, 682, 559
Showman, A. P., Cooper, C. S., Fortney, J. J., & Marley, M. S. 2008b, ApJ, 685, 1324
Showman, A. P., Fortney, J. J., Lian, Y., et al. 2009, ApJ, 699, 564
Showman, A. P., Wordsworth, R. D., Merlis, T. M., & Kaspi, Y. 2013, in Comparative Climatology of Terrestrial Planets, ed. S. J. Mackwell et al. (Tucson, AZ: Univ. Arizona Press), 277
Smagorinsky, J., Manabe, S., & Holloway, J. L. 1965, MWRv, 93, 727
Spiegel, D. S., Menou, K., & Scharf, C. A. 2009, ApJ, 691, 596
Tilley, M. A., Segura, A., Meadows, V., Hawley, S., & Davenport, J. 2019, AsBio, 19, 64
Turbet, M., Bolmont, E., Leconte, J., et al. 2018, A&A, 612, A86
Turbet, M., Leconte, J., Selsis, F., et al. 2016, A&A, 596, A112
von Paris, P., Selsis, F., Kitzmann, D., & Rauer, H. 2013, AsBio, 13, 899
Wang, J., Mawet, D., Hu, R., et al. 2018, JATIS, 4, 035001
Wang, Y., Tian, F., & Hu, Y. 2014a, ApJL, 791, L12
Wang, Y., Tian, F., & Hu, Y. 2014b, ApJL, 791, L42
Way, M. J., Aleinov, I., Amundsen, D. S., et al. 2017, ApJS, 231, 12
Williams, D. M., & Kasting, J. F. 1997, Icar, 129, 254
Williams, D. M., & Pollard, D. 2002, IJAsB, 1, 61
Williams, D. M., & Pollard, D. 2003, IJAsB, 2, 1
Wolf, E. T. 2017, ApJL, 839, L1
Wolf, E. T. 2018, ApJL, 855, L14
Wolf, E. T., Shields, A. L., Kopparapu, R. K., Haqq-Misra, J., & Toon, O. B. 2017, ApJ, 837, 107
Wolf, E. T., & Toon, O. B. 2013, AsBio, 13, 656
Wordsworth, R. 2015, ApJ, 806, 180
Wordsworth, R., Kalugina, Y., Lokshtanov, S., et al. 2017, GeoRL, 44, 665
Wordsworth, R. D., Forget, F., Selsis, F., et al. 2010, A&A, 522, A22
Yang, J., Abbot, D. S., Koll, D. D. B., Hu, Y., & Showman, A. P. 2019, ApJ, 871, 29
Yang, J., Boué, G., Fabrycky, D. C., & Abbot, D. S. 2014, ApJL, 787, L2
Yang, J., Cowan, N. B., & Abbot, D. S. 2013, ApJL, 771, L45